\documentclass[twocolumn,twoside]{revtex4}
\usepackage{graphicx}
\usepackage{fancyhdr}

\usepackage{mathtools}

\begin{document}

\title{Quarkophobic W' for LHC searches}

%

\author{A.~Gurrola$^{1}$, J.~D.~Ruiz-\'{A}lvarez$^{2}$}
\affiliation{$^{1}$Department of Physics and Astronomy, Vanderbilt University, Nashville, TN, 37235, USA \\ 
$^{2}$Instituto de Física, Universidad de Antioquia, A.A. 1226 Medellín, Colombia. }

\begin{abstract}
We consider a simplified model where a W' boson is added to the standard model with negligible couplings to quarks, but generic couplings to leptons and electroweak bosons. We study the implications of such a model for LHC searches. Consequently, we propose an LHC search through the vector boson fusion topology which would have sensitivity for such a new particle with the current proton-proton collisions's energy and available luminosity. 
\end{abstract}

\maketitle

\thispagestyle{fancy}


\section{Introduction}

In recent years, there have been several measurements that have reported deviations form the predictions of the standard model (SM) of particle physics. Experiments with different setups and experimental configurations have measured some B mesons decays with branching fractions that challenge SM theoretical predictions~\cite{BaBar:2006tnv, BaBar:2015wkg, Belle:2009zue, Belle:2016xuo, Belle:2019oag, LHCb:2019hip, LHCb:2015wdu, LHCb:2015wdu, LHCb:2014cxe}. However these measurement results have been less than 5 sigma, they have awaken the interest of the particle physics community. A plethora of models beyond the standard model have been proposed in order to explain such anomalies.

\begin{figure}[h]
\centering
\includegraphics[width=40mm]{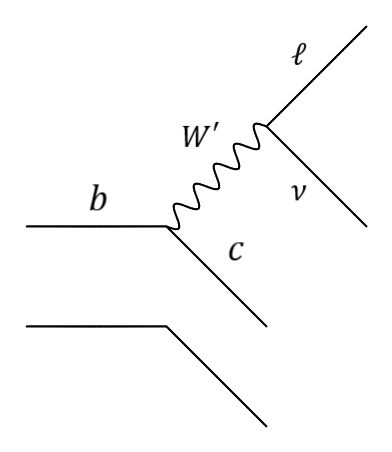}
\caption{Feynman diagram for a W' mediating a process involved in b-anomalies where the W' couples to SM quarks.} \label{ban1}
\end{figure}

It has been proved recently that these low energy anomalies should have connections with high energy physics as for proton-proton collisions at the Large Hadron Collider~\cite{Greljo:2018tzh, Fuentes-Martin:2020lea}. The most traditional model building for b-anomalies consider a new physics mediator that couples to quarks and leptons of the SM. Fig.~\ref{ban1} displays the Feynman diagram for a W' that couples to both leptons and quarks from the SM. 

\begin{figure}[h]
\centering
\includegraphics[width=40mm]{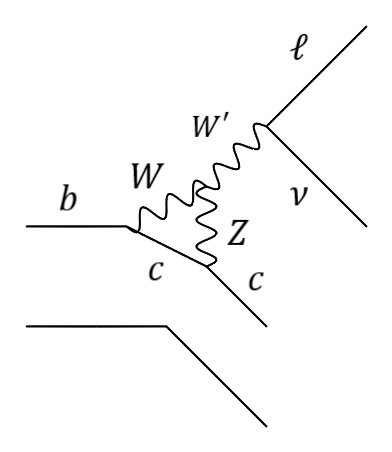}
\caption{Feynman diagram for a quarkophobic W' mediating a process involved in b-anomalies through TGC.} \label{ban2}
\end{figure}

However, one can consider a W' with negligible or no couplings to the SM quarks (quarkophobic). This W' can couple to the electroweak bosons through triple gauge couplings (TGC) which could also mediate the b-anomalies processes. The corresponding Feynman diagram for such a model is depicted in Fig.~\ref{ban2}

\begin{figure}[h]
\centering
\includegraphics[width=50mm]{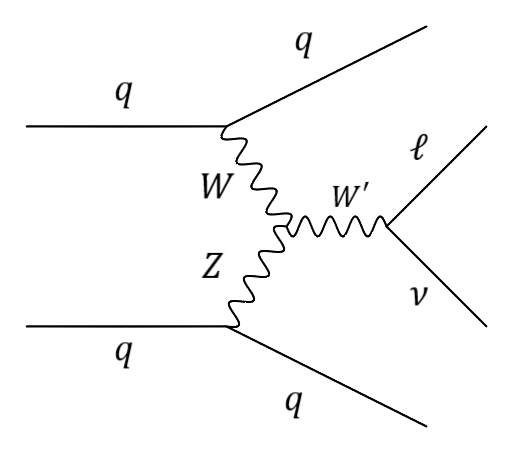}
\caption{Feynman diagram for the production of a quarkophobic W' from proton-proton collisions} \label{banVBF}
\end{figure}

A quarkophobic W' produces naturally a vector boson fusion (VBF) topology at the LHC proton-proton collisions. This fact opens the door for searching for this new particle at high energies, and additionally motivates a search for a single high energy lepton, from the W' decay, missing transverse momentum and two VBF jets. In Fig.~\ref{banVBF} is shown the VBF production of a quarkophobic W'.

We report, in a very summarized manner, a proposal for a search at the LHC of a quarkophobic W' produced through a VBF topology in proton-proton collisions at the LHC.

\section{A minimal model}

We consider to include to the SM content a new W' particle with the following minimal couplings to SM particles, as shown in Eq.~\ref{eq-1}-~\ref{eq-3}.

\begin{eqnarray}
    \mathcal{L}_{VWW'}^{1} & = & g_{1}^{V}V^{\mu}(W^{-}_{\mu\nu}W'^{+\nu}-W^{+}_{\mu\nu}W'^{-\nu} \nonumber \\
    & & +W'^{-}_{\mu\nu}W^{+\nu}-W'^{+}_{\mu\nu}W^{-\nu}),
    \label{eq-1}
\end{eqnarray}

\noindent \begin{equation}
    \mathcal{L}_{VWW'}^{2}=g_{2}^{V}(W^{+}_{\mu}W'^{-}_{\nu}V^{\mu\nu}+W'^{+}_{\mu}W^{-}_{\nu}V^{\mu\nu}),
    \label{eq-2}
\end{equation}

\noindent where $V_{\mu\nu}=\partial_{\mu}V_{\nu}-\partial_{\nu}V_{\mu}$ and $V=Z\;\text{or}\;\gamma$. The coefficients $g_{1,2}^{V}$ represent the coupling strength between the $W'$ and SM weak bosons, and thus govern the VBF production cross section. The interactions between the $W'$ and the SM leptons are described by

\begin{equation}
    \mathcal{L}_{l}=\sum_{l}\bar{\nu}_{l}\gamma_{\mu}(g^{R}_{l}(1+\gamma^{5})+g^{L}_{l}(1-\gamma^{5}))W'^{\mu}l,
    \label{eq-3}
\end{equation}

\noindent where the coefficients $g_{l}^{L,R}$ are the left-handed and right-handed couplings that govern the $W'\to l\nu$ decays. For the purpose of the studies shown in this paper, we set the values of the couplings to $\{g_{l}^{L}, g_{l}^{R}, g^{V} \} = \{ 1, 1, 1 \}$.

To have a complete model with a W' with similar couplings as the ones described, a serious amount of theoretical work must be done. However, for our purpose we only need a model that allow us to simulate the interactions of interest to have a W' produced via a VBF topology at the LHC with subsequent decays to SM leptons. In this sense, the model we propose is a simplified W' model, as the dark matter community has been doing, for LHC searches with the specificity of having TGC and negligible couplings of the W' to quarks.

\section{Samples and Strategy}
The signal model was implemented with \textsc{FeynRules}~\cite{Alloul:2013bka} and used in its UFO~\cite{Degrande:2011ua} implementation for the production of montecarlo samples. The partonic processes for signal and background events have been produced with {\textsc{MadGraph5\_aMC}} (v2.8.2)~\cite{Alwall:2014hca}. The showering and hadronization have been performed by {\textsc{PYTHIA8}}~\cite{Bierlich:2022pfr}, and the detector simulation has been done by {\textsc{Delphes}} (v3.4.2)~\cite{deFavereau:2013fsa}.

The strategy for the search is based on the characteristics of the jets coming from the VBF topology. These are, having a high pseudorapidity separation, high invariant mass and being produced in separate hemispheres of a detector like CMS or ATLAS. Additionally we have studied the kinematics of the leptons from the decay of the W' in order to further suppress SM backgrounds.

For the optimization of the selection criteria we use a signal sample with a W' with a mass of 1 TeV. As our signature is one lepton, missing transverse momentum and two jets we have considered as our main backgrounds W+jets, $t\bar{t}$ and diboson SM production. We use the significance $S/\sqrt{S+B}$ as the figure to maximize for the optimization of the selection criteria. $S$ stands for the number of signal events while $B$ represents the sum of events from all the background processes. For the proposed optimization, a dummy cross section of 0.1 pb has been used for the W'. Real cross sections can be easily obtained from MadGraph accordingly to specific values of the couplings and W' mass.

\section{Results}

We describe in the following the results for the case where the W' decays into a muon. We have also studied the electron and tau channels, and these results will be put together in a complete separate paper. 

\begin{figure*}[t]
\centering
\includegraphics[width=70mm]{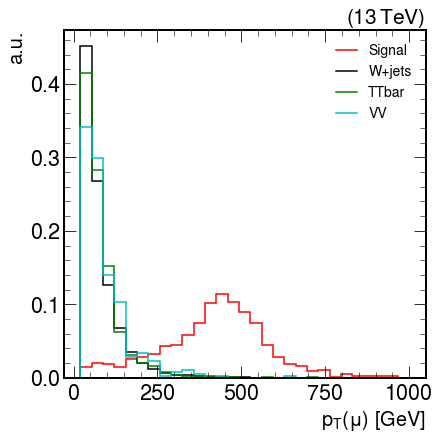}
\includegraphics[width=70mm]{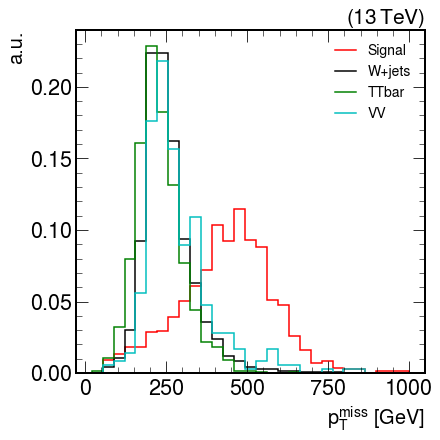}
\includegraphics[width=70mm]{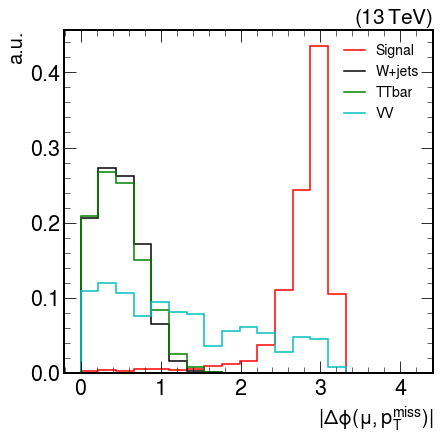}
\includegraphics[width=70mm]{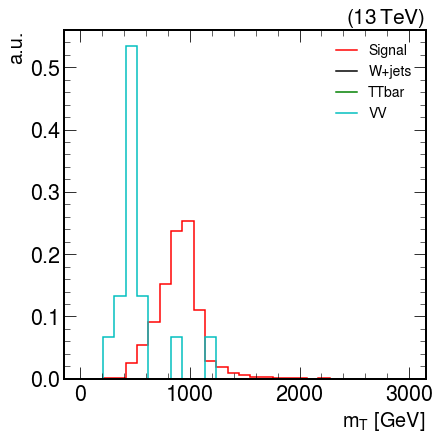}
\caption{Signal and backgrounds normalized to unit distributions for (top-left) $p_{T}$ of the muon, (top-right) $p_{T}^{miss}$, (bottom-left) the absolute value of the $\Delta\phi$ betwwen the muon and the missing transverse momentum and (bottom-right) the transverse mass of the muon and the missing transverse momentum} \label{mucha}
\end{figure*}

In Fig.~\ref{mucha} are shown the characteristics of the muon and the missing transverse momentum that were found to be useful to discriminate signal from backgrounds. Additionally, the transverse mass distribution for the muon is shown. For a complete search this last distribution could be considered as the variable of interest. 

The selection criteria developed is:
\begin{enumerate}
    \item A baseline selection consisting of requiring at least two jets with $p_{T}>60$ GeV, $N(b)=0$, $N(\mu)=1$, and veto leptons with $p_{T}(l)>25$ GeV
    \item $\eta(j_{1})\times\eta(j_{2})<0$
    \item $m_{jj}>1000$ GeV
    \item $|\Delta(\eta(j_{1}),\eta(j_{2}))|>4.0$
    \item $p_{T}(\mu)>200$ GeV
    \item $p_{T}^{miss}>200$ GeV
    \item $|\Delta(\phi(\mu),p_{T}^{miss})|>1.0$
\end{enumerate}

In Table~\ref{table:MuEvyields} are shown the event yields for the signal with $m(W')=1$~TeV and the SM backgrounds at each step of the selection. Also is shown the significance as previously defined for each selection criteria. The number of events are calculated for proton-proton collisions at the LHC at a centrer of mass energy of 13 TeV and a luminosity of 150~$\text{fb}^{-1}$.

\begin{table}[h]
\begin{center}
\caption{Event yields for muon channel signal and backgrounds at each selection step.}
\begin{tabular}{p{0.1\linewidth}*{6}{c}}

\  & Signal & W+jets & $t\bar{t}$ & VV & $S/\sqrt{S+B}$\\ 
\hline 
Initial & 1259.4 & 152156.8 & 48246.6 & 1870.3 & 2.8 \\ 
Cut 1 & 930.3 & 68480.3 & 14364.6 & 788.2 & 3.2 \\ 
Cut 2 & 593.7 & 13461.8 & 1796.9 & 147.8 & 4.7 \\ 
Cut 3 & 527.1 & 2771.8 & 365.0 & 27.9 & 8.7 \\ 
Cut 4 & 474.0 & 118.4 & 17.6 & 2.1 & 19.2 \\ 
Cut 5 & 459.9 & 91.7 & 11.6 & 1.9 & 19.3 \\ 
Cut 6 & 459.6 & 0.0 & 0.0 & 0.9 & 21.4 \\
\hline

\end{tabular}
\label{table:MuEvyields}
\end{center}
\end{table}

\section{Conclusions}

We have proposed a model for searches at the LHC looking for a W' produced through the VBF topology. We have also developed a search that is capable of efficiently looking for such a W' in the context of the data collected at the LHC for the run 2. We have shown the details for this search when the W' decays into a muon a neutrino but we will be producing a complete paper were also the tau and electron channels are included. Our finding strongly motivates for searching for a heavy quarkophobic W' with the current data at the LHC.

\bigskip 
\begin{acknowledgments}
A.G. gratefully acknowledge the constant support of the Physics \& Astronomy department at Vanderbilt University and the US National Science Foundation. This work is supported in part by NSF Award PHY-1945366. J. D. R.-\'{A}. gratefully acknowledges the support of Universidad de Antioquia, the Colombian Science Ministry and Sostenibilidad-UdeA.
\end{acknowledgments}

\bigskip 

\end{document}